\documentclass[aps,twocolumn,showpacs,preprintnumbers,amsmath,amssymb,floatfix]{revtex4-1}

\usepackage{graphicx,tabularx}
\usepackage{dcolumn}
\usepackage{bm}
\usepackage{color}
\usepackage{epstopdf}
\usepackage{upgreek}
\usepackage{color}

\begin{document}

\title{High-field and high-temperature magnetoresistance reveals the superconducting behaviour  of the stacking faults in multilayer
graphene}


\author{Christian E. Precker$^1$}
\affiliation{Division of Superconductivity and Magnetism, Felix-Bloch Institute for Solid-state
  Physics, University of Leipzig, 04103 Leipzig, Germany}
\author{Jos\'e Barzola-Quiquia$^2$}
\affiliation{Division of Superconductivity and Magnetism, Felix-Bloch Institute for Solid-state
  Physics, University of Leipzig, 04103 Leipzig, Germany}
\author{Mun K. Chan}
\affiliation{National High Magnetic Field Laboratory, Los Alamos National Laboratory, Pulsed Field Facility, Los Alamos, New Mexico 87545, USA.}
\author{Marcelo Jaime}
\affiliation{National High Magnetic Field Laboratory, Los Alamos National Laboratory, Pulsed Field Facility, Los Alamos, New Mexico 87545, USA.}
\author{Pablo D. Esquinazi}
\affiliation{Division of Superconductivity and Magnetism, Felix-Bloch Institute for Solid-state
  Physics, University of Leipzig, 04103 Leipzig, Germany}



\begin{abstract}
In spite of  40 years of experimental studies and several theoretical proposals, an overall 
interpretation of the  complex 
behavior of the magnetoresistance (MR) of multilayer graphene, i.e. graphite, at high fields ($B \lesssim 70~$T) and in a broad temperature range 
is still lacking.  Part of the complexity is due to the contribution of 
 stacking faults (SFs), which most of  thick enough multilayer graphene samples have. We propose 
a  procedure that  allows us to extract the SF contribution to the MR we have 
measured  at 0.48~K $\leq T \leq$ 250~K 
and  0~T$\leq B  \lesssim$ 65~T. We found that the MR behavior of part of the SFs 
is similar to that of granular superconductors with a superconducting  
 critical temperature $T_c \sim $ 350~K, in agreement with recent publications. The
measurements were done on a multilayer graphene TEM lamella, contacting the edges of the two-dimensional 
SFs. 

\end{abstract}

\maketitle

\section{Introduction}
\label{introduction}

The recent discovery of superconductivity in twisted bilayer graphene \cite{cao18,yan19}, a stacking fault in itself, 
in trilayer graphene moir\'e superlattice  \cite{che19} as well as in rhombohedral stacking order  \cite{zho21} at $T < 5~$K  supports the 
assumption that  the origin of the ``hidden superconductivity" reported in several bulk and mesoscopic graphite 
samples in the last 50 years \cite{ant74,ant75,yakovjltp00,esq08,esqpip},
is related to the existence of  two-dimensional stacking faults (SFs). Because these SFs are embedded in a 
multilayer-graphene matrix with a Bernal (2H) or rhombohedral (3R) stacking order, they can play a main role in
the measured transport properties.   
Previous
studies demonstrated  that SFs, like  twisted graphene layers, are common in 
 well-ordered graphite samples  \cite{kuw90,mil10,flo13}. 
Therefore,  flat bands regions are expected to be found  in bulk and mesoscopic graphite samples 
at certain SFs,  where  superconductivity at low and  high temperatures is predicted  
 \cite{kop13,mun13,hei16,vol18,cea19}.  
 
The tuning of superconductivity in mesoscopic bi- and trilayers graphene samples through the fabrication of regions with well-defined twist angle,
added to the possibility of contacting directly the superconducting region, are clear advantages with respect to experimental studies
in bulk or mesoscopic graphite samples. Although the existence of a large number of SFs can be recognized by Scanning Transmission Electron Microscopy (STEM) and X-rays Diffraction (XRD) analysis \cite{Esquinazi2017}, the twist angle distribution remains  largely  unknown in thick multilayer graphene samples. 
However, there are  some advantages in the study of the behavior of the SFs in well-ordered multilayer graphene or 
graphite samples. One of them is that  SFs  of very large 
areas (several 100's of $\mu$m$^2$) with high degree of lattice order  can be found well shielded from environmental influence. 
Another advantage is that   thick enough samples can  have up to $\sim 25\%$  of  the 3R stacking, in addition to the main 2H-type stacking \cite{kelly,wu22}.
The  SFs between untwisted 2H and 3R crystalline regions are expected to have high  superconducting temperatures \cite{kop13,mun13,hei16,vol18}. 
This kind of SF has not yet been produced in a controllable way. On the other hand, if superconductivity is located at certain regions of such large SF areas,   
we expect to have  granular and not homogeneous superconductivity, i.e. Josephson coupled superconducting regions of several tens of  $\mu$m$^2$ each. 
Our  research covers  the study of the low and very high field behavior of the magnetoresistance (MR) of the SFs in multilayer graphene extracted
from the measured MR with the help of a procedure we developed. The results clarify several open questions  on the interpretation of the MR of
graphite and indicate the existence of superconducting regions with an upper critical temperature of $\sim 350~$K.

\section{Methods and sample}
We have prepared a multilayer graphene TEM lamella 
of dimensions $20~\mu$m $\times 5~\mu$m $\times 0.6~\mu$m, with the width  in the $c$-axis direction, obtained from a highly oriented pyrolytic graphite (HOPG) bulk sample using a similar procedure as in \cite{bal13},  and investigated the transport properties at low- and high-fields (applied parallel to the $c$-axis) 
in a broad  temperature range.  The lamella was fixed on a substrate combining electron beam 
lithography with SiN$_x$ deposition covering part of the sample surface.
Afterwards, the lamella was 
 inductively etched with a plasma reactive ion etching system (ICP-RIE) 
to take out the disordered graphite layer formed during the milling process.
Four electrodes were prepared with electron beam lithography and depositing  Cr/Au, see the sample optical image in the inset of Fig.\ref{Fig:Fig2}(b). The sample electrical resistance at 300~K was 7.44~$\Omega$. As we show below, such a TEM lamella with the graphite $c-$axis parallel to the substrate
provides the best way to get the SFs contribution to the total electrical resistance by 
contacting their edges. Considering the width (in the $c-$axis direction) of the sample and the
corresponding STEM images in samples from the same batch, see Refs. \cite{bal13,Esquinazi2017}, the number
of SFs is significant. Therefore,  our electrical voltage contacts pick up the response of 
several of them; the current input is distributed through all SFs. In this way, we expect to
get the superconducting response of the SFs with the highest critical temperature.

For measurements at  2~K $\leq T \leq $ 300~K and DC magnetic fields $B \leq 7$~T  we  used
 a $^4$He cryostat, prior to the high field pulsed measurements. 
These last  measurements were performed at the National High Magnetic Field Laboratory's Pulsed Field Facility (NHMFL-PFF) at Los Alamos National Laboratory (LANL) \cite{Jaime2006}. The measurements were done in a cryostat with a temperature range  $T$ = 0.45 ~K to 250~K, equipped with a 65~T multi-shot magnet, powered by a 32 mF, 4 MJ capacitor bank with a pulse duration of $\sim 70$~ms \cite{Swenson2004, Nguyen2016}. Most of the experiments were performed with pulses of 60~T.  An alternating current of 12~$\mu$A amplitude was applied to the sample at a frequency of 50.5 kHz. The voltage was measured with a 20~MHz sampling rate. The field was always 
 normal to the graphene planes and the SFs.

\section{Results and discussion}

 Electrical transport measurements under high magnetic fields ($B > 10~$T) performed in bulk  and millimeter long multilayer graphene samples were reported in the last 40 years \cite{Tanuma1981,Yaguchi1998,Fauque2013,Akiba2015}. 
The observed behavior of the MR of those samples is complex and non-monotonous in field. Several interpretations were proposed, namely: fluctuations of charge density waves \cite{Yoshioka1981, Timp1983}, magnetic freeze-out of carriers \cite{Brandt1974}, 3D quantum Hall effect through the appearance of chiral surface states \cite{Fauque2013, Bernevig2007}, the emergence of an excitonic BCS-like state \cite{Akiba2015}, the appearance of an insulating surface states that carry no charge or spin within the planes \cite{Arnold2017}, magnetic catalysis scenario~\cite{GORBAR-02,KHV-01,DETAR-16}, to cite a few of them. However, none of those studies considered the
parallel contributions of at least two subsystems in graphite \cite{gar12,zor17}, i.e., the MR of the SFs and the one from the graphite matrix with mostly Bernal
stacking order. 

The incorrect interpretations of the transport, as well as the magnetization properties of graphite found in several early reports, relied
on the assumption of electrically homogeneous samples. 
The lack of  TEM or STEM characterization with the electron beam parallel to the
graphene planes necessary to get an evidence of the existence of SFs, impeded in the past a
timely development of the physics of graphite and of its SFs. One prominent example is the common, incorrect assumption
that graphite is a semimetal with a finite Fermi surface at low temperatures.  Systematic transport studies 
as a function of the thickness of graphite samples proved, however, that 
 the SFs   substantially contribute to the electrical transport and 
 are at the origin of Shubnikov-de Haas (SdH) (or de Haas-van Alphen in the magnetization) 
 quantum oscillations \cite{zor18}. The vanishing of the SdH oscillations amplitude the smaller the thickness
 of the graphite samples, maintaining their high structural ordering, was recognized already at the beginning of 2000  without providing
 a clear interpretation  of this behavior \cite{oha00,oha01}. The change in the temperature and field dependence of the electrical 
 transport as a function of the graphite sample thickness  already indicated that
 the metalliclike behavior vanishes when the thickness of the sample is smaller than the average distance between the SFs \cite{bar08},
 which for the HOPG sample we used in this study means a length in the $c-$axis direction of less than $\sim 30~$nm. From
 different characterizations \cite{gar12,esq14,zor17,Barzola2019,reg21} we know nowadays that the intrinsic properties of {\em ideal graphite}, i.e.
 without SFs,  are compatible
 with those of a narrow band semiconductor, not a semimetal and a finite Fermi surface does not exist at low temperatures. 
 Stacking faults with superconducting behavior can be also found 
  in  mesoscopic and bulk samples, see  \cite{Esquinazi2017} and Refs. therein.

\begin{figure}
\centering
\includegraphics[width=\columnwidth]{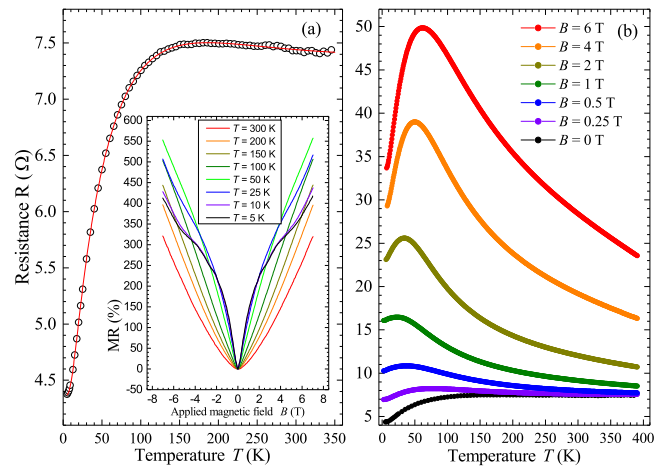}
\caption{\label{Fig:Fig1} (a) Temperature dependence of the electrical resistance of the TEM lamella at $B=0$~T. 
The open circles represent the experimental data and the red line a fit to Eq.~(\ref{Eq:RT}). The inset shows the MR at different constant temperatures at fields between $ B \pm 7$~T. (b) Temperature dependence of the resistance at different magnetic fields.}
\end{figure}

The temperature dependence of the resistance at $B = 0$ is shown in Fig.\ref{Fig:Fig1}(a). The resistance $R(T)$ increases 
with temperature to $T \simeq 165~$K,  decreasing at higher $T$. This  is one possible behavior of $R(T)$ 
for bulk and thick flakes of graphite reported in the  literature \cite{gar12,zor17}. The red line in Fig.\ref{Fig:Fig1}~(a) is a  fit to a parallel resistance model 
 given  by the contribution of the SFs $R_\textrm{i}(T)$ and of the main ideal graphite matrix $R_\textrm{2H}(T)$ \cite{gar12}:
\begin{equation}
\label{Eq:RT}
	\dfrac{1}{R(T)} \simeq \dfrac{1}{R_\textrm{i}(T)} + \dfrac{1}{R_\textrm{2H}(T)},
\end{equation}
where  $R$ is the total sample resistance (see Refs.~\cite{gar12,zor17} for the full expression).  
$R_\textrm{i}(T)$  has an  increasing with temperature contribution  given by an exponential function
of the form
	$R_\textrm{i}(T) \propto  \exp(-E_\textrm{a}/k_\textrm{B} T)$. The thermal activation energy $E_\textrm{a}  \simeq 3$~meV is obtained from the fit. 
	We note that  this temperature dependent behavior  
is expected for  granular superconductors according to Refs. \cite{sha83,lan67,mcc70,str07}.  

The narrow-gap semiconducting contribution of the 2H matrix can be approximated as 
	$R_\textrm{2H}(T) \propto  \exp(E_{g,\textrm{2H}}/(2k_\textrm{B} T)$ with an energy gap $E_{g,\textrm{2H}} \simeq 52$~meV obtained from the fit. All fit parameters
	are similar to  those reported in the literature \cite{zor17,gar12}, emphasizing that the measured sample is representative.  Due to its
small amount, we neglect the semiconducting parallel contribution of the 3R matrix to the total $R(T)$. We note that the excellent fits of $R(T)$ 
obtained for graphite samples of different thickness  using the parallel resistance model and those exponential terms is not obtained by replacing them with  $T^n$  or
other dependencies found in the literature \cite{zor17}. 
	
	The inset in Fig.\ref{Fig:Fig1}(a) shows the magnetoresistance  defined as
	MR $ = (R(B)-R(0))/R(0)$
measured at different constant temperatures with magnetic fields between $ \pm 7$~T. At $T \leq 25$~K,  SdH oscillations are detected with the main  period in $B^{-1}$  of 0.21~T$^{-1}$ in agreement  with the literature. From this period we estimate a 2D carrier density  $n_\textrm{2D} \simeq 2.3 \times 10^{11}$~cm$^{-2}$ at certain  SFs that originate the SdH oscillations \cite{zor18}. Fig.\ref{Fig:Fig1}(b) shows the $T-$dependence of the electrical resistance of the lamella at different magnetic fields. The sample shows the typical re-entrant metallic behavior at $B \gtrsim 1~$T and $T < 100~$K reported for bulk graphite samples \cite{yakovprl03}.

\begin{figure}
\centering
\includegraphics[width=\columnwidth]{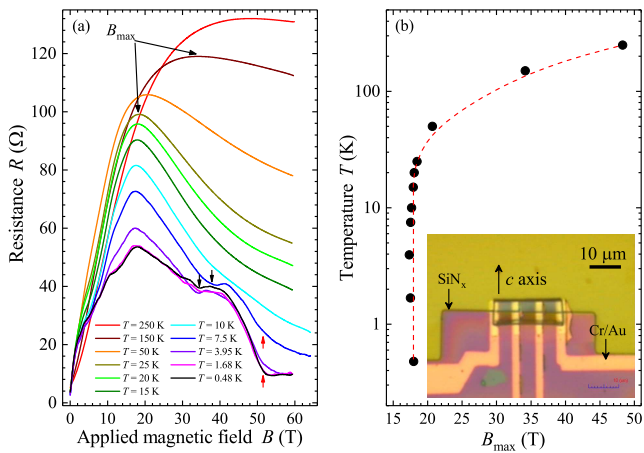}
\caption{\label{Fig:Fig2} (a) Pulsed magnetic field dependence of the resistance at different constant temperatures. The arrows  point 
out the fields at the  electronic phase transitions in graphite observed at $T < 10~$K. (b) Field $B_\textrm{max}$ where the MR has its
maximum (see (a)) vs temperature. The red dashed line is a guide to the eye. The inset shows an optical image of the
sample with its contacts and substrate.}
\end{figure}

Fig.\ref{Fig:Fig2}(a) shows the absolute value of the electrical resistance of the lamella vs.  magnetic field at different  temperatures. 
The electronic transitions $\alpha$ ($\downarrow$) and $\alpha '$ ($\uparrow$) are clearly recognized at $T < 10$~K in the field range 30~T~$< B <$~55~T.  
Fig.\ref{Fig:Fig2}(b) shows the $T-$dependence of the  field  $B_\textrm{max}$ at which the MR shows a maximum, see Fig.\ref{Fig:Fig2}(a). 
We found  that $B_\textrm{max}$ remains $T-$independent at  $T \lesssim 25$~K. Above this temperature $B_\textrm{max}$ increases. This behavior is in very good  agreement with that reported several times in the last 40 years \cite{Tanuma1981,Yaguchi1998,Iye1982,Ochimizu1992,Akiba2015,Arnold2017}. In particular, the behavior of $B_\textrm{max}(T)$ was explained  on the basis of  the magnetic catalysis model \cite{KHV-01}. From all the measured data we conclude that this 
TEM lamella shows all the characteristics of the electrical resistance and magnetoresistance of 
bulk graphite samples. We provide below an  interpretation of the observed behavior. 

We will extract  first the SF  resistance ${R_\textrm{i}(T,B)}$ from the measured $R(T,B)$ data. For that we   rewrite Eq.(\ref{Eq:RT}) at a constant temperature as:
\begin{equation}
\label{Eq:Ri_B}
	R_\textrm{i}(B) = {R(B)}/[{1 - \left({R(B)}/{R_\textrm{2H}(B))}\right]}.
\end{equation}
The MR of the  2H contribution can be described by the two-band model (TBM) appropriate for semiconductors and  derived under the Boltzmann-Drude quasi-classical diffusive approach \cite{kelly}.  As emphasized above, the transport properties of ideal graphite (without SFs) match the ones of a narrow-band 
gap semiconductor. Therefore, we approximate the field-dependent resistance related to the 
semiconducting
2H-contribution  as:
\begin{equation}
	\label{Eq:R2H_TBM}
	R_\textrm{2H}(B) \simeq R_\textrm{2H}(0) \cdot \left[1 + \dfrac{\mu^2B^2 \left( 1 - \dfrac{\Delta n^2}{n^2}\right)}{1 + \mu^2B^2 \left( \dfrac{\Delta n^2}{n^2} \right)} \right],
\end{equation}
where we have assumed equal mobility for both electrons and holes  ($\mu = \mu_e \approx \mu_h$), and $\Delta n / n = (n_e - n_h)/(n_e + n_h)$ is the relative charge imbalance between electron $n_e$ and hole $n_h$ carrier densities.   

The simplified expression of Eq.(\ref{Eq:R2H_TBM}) has only two adjustable fitting parameters: the average mobility $\mu$ and the relative charge imbalance $\Delta n / n$; $R_\textrm{2H}(0)$ is a fixed parameter obtained from the fit in Fig.\ref{Fig:Fig1}(a). Eq.(\ref{Eq:R2H_TBM}) provides two key features of the MR of the semiconducting contribution, namely, the $B^2$ field dependence at low fields and its saturation at high enough fields. 

\begin{figure}
\centering
\includegraphics[width=\columnwidth]{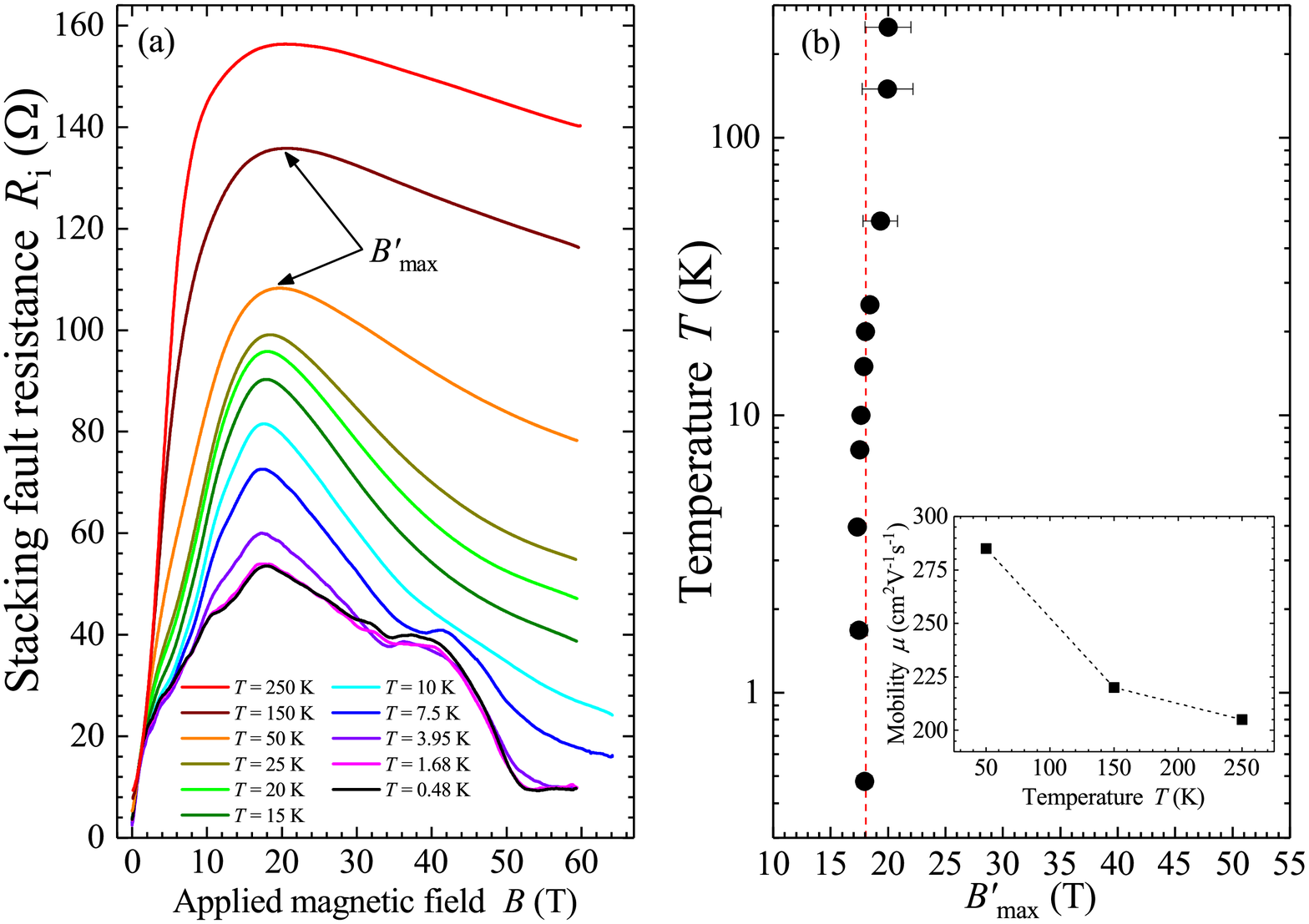}
\caption{\label{Fig:Fig3} (a) Field dependence of the SF  resistance $R_\textrm{i}(B)$, calculated using Eq.(\ref{Eq:Ri_B}), at different constant temperatures. (b) $B'_\textrm{max}$ vs temperature obtained from (a). The inset shows  the temperature dependence of the carrier mobility.}
\end{figure}

Replacing Eq.(\ref{Eq:R2H_TBM}) in  Eq.(\ref{Eq:Ri_B}), we obtain  $R_\textrm{i}(B)$ plotted in Fig.\ref{Fig:Fig3}(a).
The results indicate that  at $T < 25$~K,  $R(B) \simeq R_i(B)$ 
because the semiconducting contribution becomes negligible, i.e. $(R_\textrm{2H}(B,T)/R(B,T))|_{T < 25\textrm{K}} \gg 1$. 
We further note that  $R_\textrm{i}(B)$  shows a maximum at  $B'_\textrm{max}  \simeq 18$~T, which does not depend significantly on $T$ within error, see Fig.\ref{Fig:Fig3}(b).   
These results 
  indicate 
 that the temperature shift  of $B_\textrm{max}(T)$ in the MR, see Fig.\ref{Fig:Fig2}(b),  is an artifact caused by the growing influence  at  
$T > 25$~K of the  semiconducting contribution $R_\textrm{2H}(B)$ in parallel to the SFs one. 
Regarding the parameters used, the charge imbalance between electrons and holes was considered constant $\Delta n / n = 0.05$, and the 
obtained mobility $\mu (T)$ decreases with  temperature (see  inset in  Fig.\ref{Fig:Fig3}(b)), in   qualitative agreement with the behavior found in the literature \cite{Pendrys1980,esq12}. 

The MR of the SF ($R_\textrm{i}(B)$) plotted in  Fig.\ref{Fig:Fig3}(a), resembles the one observed in  granular superconductors, 
like granular Al in a Ge matrix or InO films \cite{Shapira1983,Gerber1997,Gantmakher1996}. In particular, it shows a linear increase with field 
at low fields and decreases at fields above a certain field. 
The explanation for the linear increase with field discussed in the literature is based on the influence of the field in 
the Josephson coupling between superconducting regions or in our case 2D regions (or `grains") at some SFs. 
The higher the field, the larger is the number of uncoupled
superconducting regions  and the resistance increases linearly.  
After a  maximum number of independent regions is reached at $B \sim B'_\textrm{max}$, a higher field 
 increases the density of states  inside  those regions, increasing the probability of having 
 Cooper pairs and  
the resistance starts to decrease with field. In this field range, the intragrain superconducting fluctuations affect the intergrain conductivity reducing  the total resistance. 
This appears  to be a general behavior in granular superconductors, see Fig.~7 in Ref.~\cite{Barzola2019} and Refs. therein. 
 We expect therefore that  the field at which $R_\textrm{i}(B)$ starts to saturate can  be  considered as a critical field $B_\textrm{c2}$. 
This appears to be the case at a field $B \simeq B_{c2}  \sim 50~$T $\sim 3 B'_\textrm{max}$  at $T < 10~$K, see Fig.\ref{Fig:Fig3}(a). 
However, we expect  that  $B_\textrm{c2}$ decreases with temperature, which is not clearly observed in $R_\textrm{i}(B)$ of Fig.\ref{Fig:Fig3}(a) at $T \gtrsim 10~$K. 
The absence of a clear saturation at high temperatures and fields could be due to  superconducting fluctuations, 
which in granular superconductors are expected to persist up to very high fields and temperatures \cite{Gantmakher1996}. 

\begin{figure}
\centering
\includegraphics[width=\columnwidth]{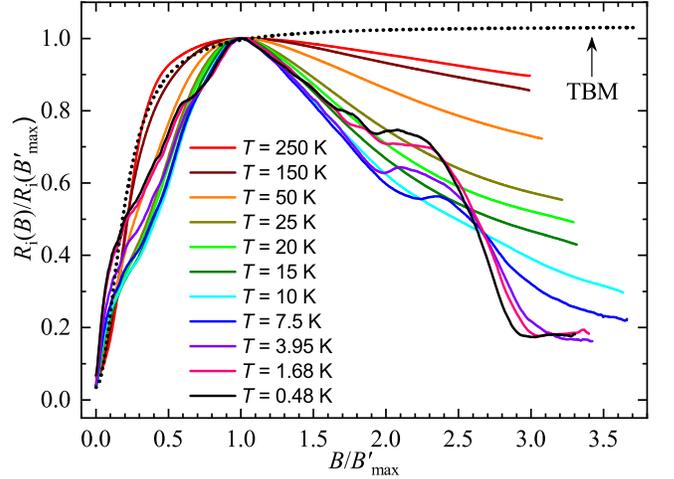}
\caption{\label{Fig:Fig4} Normalized SF contribution $R'_\textrm{i}$ vs normalized magnetic field at different constant temperatures. 
The dotted curve is the normalized MR obtained using the TBM, see  Eq.~(\ref{Eq:R2H_TBM}).}
\end{figure}

Fig.\ref{Fig:Fig4} shows the normalized $R_\textrm{i}(B)$ vs the normalized magnetic field at different temperatures. We note that the higher the temperature 
the smaller is the  decrease of the resistance with field at  $B > B'_\textrm{max}$. This is expected because 
the number and/or size of the superconducting regions inside the SFs
 should in this case decrease.
Therefore,  at the highest critical temperature 
$T_\textrm{c}$ of the superconducting grains, there should not be a decrease with field of $R_\textrm{i}(B)$. At $T \geq T_c$  we
expect a MR behavior similar to the semiconducting matrix, approximately given by the TBM, see  Fig.\ref{Fig:Fig4}.   

Together with the similarities of our results  to those of granular superconductors, let us emphasize here 
why we expect to have granular superconductivity at certain SFs and not a homogeneous state. Granular superconductivity 
occurs because the flat bands formed at the 2D SFs  are not homogeneous in areas more than a few tens of micrometer square. 
This is obvious if we take into account the STEM evidence about the order or disorder that usual graphite  samples have. 
Not only the perfection of a 2D SFs in the corresponding plane is an issue but also, e.g., the flat bands can be  
affected by the number of ideal graphene layers on both side of the interface. 
In addition, the graphene layers are not ideal  over the sample, but have boundaries that restrict  the
homogeneous regions. This is a fact that is simple to recognize from STEM images taken at energies $\sim 30~$keV. 
The granular nature was already shown to be highly likely in several reports, as for example \cite{bal13,schcar,bal15}. 
At low enough temperatures and currents in the nA region, $I-V$  curves indicate indeed a Josephson
behavior and zero resistance within error \cite{bal13}. Even the reported transition in twisted bilayer graphene mesoscopic samples 
does not appear to behave as a homogeneous but as a granular superconductor, as a direct, quantitative comparison between those 
results and the ones obtained in graphite TEM lamellae indicates \cite{esq19}.

\begin{figure}
\centering
\includegraphics[width=\columnwidth]{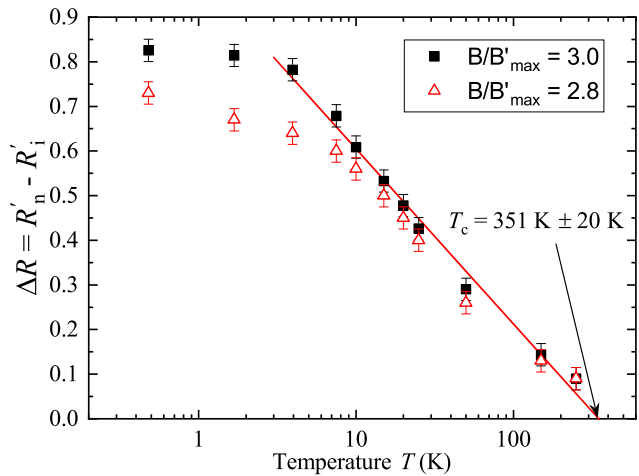}
\caption{\label{Fig:Fig5} Unitless difference between the normalized normal state and SF resistances at $B/B'_\textrm{max} = 3$ and 2.8 vs. temperature.}
\end{figure}

The proposed interpretation of the MR of the SFs in terms of Josephson-coupled superconducting regions  at certain interfaces or SFs,
implies that at a fixed field the   temperature dependence of the resistance should be compatible with the one expected for
 2D granular superconductors. An analytical expression for the  resistance of this 2D system within the 
 effective medium approximation has been obtained in \cite{str07}. In particular, at  fields near the critical field or at 
 high enough temperatures the resistance between superconducting grains reaches a critical resistance $R_{JC}(T)$, which self-consistent 
 solution (see Fig.3 in \cite{str07}) follows nearly a $\ln(T/T_c)$ at $T/T_c > 0.2$, independently of the value of the assumed 
 charging energy. We  compare  qualitatively this prediction with  the difference    between the normalized resistance in the normal state 
 $R'_\textrm{n}$ and the normalized SF resistance ($R_\textrm{i}^{'} = R_\textrm{i}(B)/R_\textrm{i}(B'_\textrm{max})$)
from  Fig.\ref{Fig:Fig4}, $\Delta R = R'_\textrm{n} - R_\textrm{i}^{'}$ at $B/B'_\textrm{max} = 2.8$ and 3. The difference $\Delta R$ follows a $ \sim \ln(T/T_\textrm{c})$ at high
enough temperatures and a critical temperature $T_\textrm{c} = (351 \pm 20)$~K is obtained by extrapolation to $\Delta R = 0$, see  Fig.\ref{Fig:Fig5}. 
Interestingly, this $T_\textrm{c}$ agrees with the one 
suggested by different transport, magnetization \cite{Precker2016,lay22,rou22} and magnetic force microscopy \cite{sti18,ari22} measurements in  different, well ordered 
natural graphite samples.
 At lower $T$, the  behavior of $\Delta R(T)$ is affected by the transition to the normal state or by the electronic
transitions, see Fig.\ref{Fig:Fig4}, preventing a comparison with the predicted $R_{JC}(T)$ in the whole temperature range.

\section{Conclusion} The MR of a multilayer graphene TEM lamella  shows  a temperature-dependent maximum at $B_\textrm{max}(T)$, which increases
with temperature in agreement with earlier measurements of large graphite samples. 
Assuming that the MR is given by the parallel contribution of a semiconducting
graphite matrix and of the stacking faults, we were able to extract the MR of this last in a broad temperature and magnetic field ranges. 
Our results indicate that the observed temperature dependence of $B_\textrm{max}(T)$ is an artifact due to the 
increasing contribution of  
the semiconducting graphite matrix with temperature. The extracted  stacking fault MR shows several features compatible with those found 
in granular superconductors. The extrapolated maximum 
superconducting critical temperature of $\sim 350~$K for the superconducting regions at the
stacking faults is in agreement with recent reports.

Acknowledgments: The authors thanks Zhipeng Zhang for the deposition of SiN$_x$, and Marius Grundmann for giving access the to ICP-RIE. C.E.P. gratefully acknowledges the support provided by the Brazilian National Council for the Improvement of Higher Education (CAPES) under 99999.013188/2013-05. MJ and MKC acknowledge support by the US DOE Basic Energy Science project ``Science at 100T". The studies were partially supported by the DAAD Nr. 57207627, by the DFG under ES 86/29-1 and  31047526 (SFB 762) and the European Regional Development Fund Grant Nr.: 231301388.  A portion of this work was performed at the National High Magnetic Field Laboratory, which is supported by the NSF Cooperative Agreement No. DMR-1644779, the US DOE and the State of Florida.

Data Archival: All the data included in the figures will be available at https://speicherwolke.uni-leipzig.de/index.php/s/X3TPRBwHYKrJM54 and on
request from the corresponding author (PDE).


\noindent
$^1${ceprecker@gmail.com; Current address: AIMEN Technology Centre, Smart Systems and Smart Manufacturing-Artificial Intelligence and Data Analytics Laboratory, PI. Cataboi, 36418 Pontevedra, Spain}\\
$^2${Current address: Goldschmidtstrasse 21, 04103 Leipzig, Germany}

\end{document}